\def\be{\begin{equation}}
\def\ee{\end{equation}}
\def\ba{\begin{array}}
\def\ea{\end{array}}

\documentclass[prl,showpacs,twocolumn,amsmath]{revtex4}
\usepackage{amsfonts}
\usepackage{mathrsfs}
\usepackage{amssymb}
\usepackage{pifont}
\usepackage{epsfig,subfigure,dsfont,amsthm,amsbsy,mathrsfs,amscd}
\def\qed{\leavevmode\unskip\penalty9999 \hbox{}\nobreak\hfill
     \quad\hbox{\leavevmode  \hbox to.77778em{%
               \hfil\vrule   \vbox to.675em%
               {\hrule width.6em\vfil\hrule}\vrule\hfil}}
     \par\vskip3pt}

\begin{document}
\title{\large\bf Hawking effect can generate physically inaccessible genuine tripartite nonlocality}

\author{ Tinggui Zhang$^{1, \dag}$ Xin Wang$^{1}$ and Shao-Ming Fei$^{2,3}$}
\affiliation{ ${1}$ School of Mathematics and Statistics, Hainan Normal University, Haikou, 571158, China \\
$2$ School of Mathematical Sciences, Capital Normal University, Beijing 100048, China \\
$3$ Max-Planck-Institute for Mathematics in the Sciences, Leipzig 04103, Germany\\
$^{\dag}$ Correspondence to tinggui333@163.com}

\bigskip

\bigskip
\begin{abstract}
We explore the acceleration effect on the genuine tripartite nonlocality
(GTN) for one or two accelerated detector(s) coupled to the vacuum field
with initial mixed tripartite states. We show that the Hawking radiation degrades the physically accessible GTN, which suffers from ``sudden death" at certain critical Hawking
temperature. An novel phenomenon has been observed first time that the Hawking
effect can generate the physically inaccessible GTN for fermion
fields in curved spacetime, the ``sudden birth" of the physically inaccessible GTN.
This result shows that the GTN can pass through the event horizon of black hole
for certain mixed initial states. We also derived analytically the tradeoff relations of genuine tripartite entanglement (GTE) and quantum coherence under the influence of
Hawking effect.
\end{abstract}

 \maketitle

\section{Introduction}
Raised by Einstein et al. \cite{abnr} in discussions on the
incompleteness of quantum mechanics in 1935, quantum nonlocality
\cite{hczg} has been extensively studied and played significant
roles in a variety of quantum tasks \cite{ndsv}. Recent studies have
shown that quantum nonlocality is also a unique quantum resource for
some device-independent quantum tasks such as the key distribution
\cite{aanb}, random expansion \cite{spaa}, random amplification
\cite{rcol} and the related experiments \cite{mxws,wlml,lkyz}. In
particular, in 1987 a special kind of nonlocality called genuine tripartite nonlocality (GTN) was introduced by Svetlichny for tripartite systems \cite{gsve}.
Such multipartite nonlocal correlations have not only foundational
implications \cite{ngis} but also novel applications in quantum
communication and quantum computation \cite{msju,msee,pzet,cylu}, as
well as in phase transitions and criticality in many-body systems
\cite{pzet}. GTN can be detected by the violation of
Svetlichny inequality. Let $A_i=\vec{a}_i\cdot\vec{\sigma}$,
$B_i=\vec{b}_i\cdot\vec{\sigma}$ and
$C_i=\vec{c}_i\cdot\vec{\sigma}$, $i=0,1$, be the measurement
observables associated with the first, second and third qubit, respectively, with
$\vec{a}_i$, $\vec{b}_i$ and $\vec{c}_i$ the real unit vectors,
$\vec{\sigma}=(\sigma_1,\sigma_2,\sigma_3)$ the vector given by the standard
Pauli matrices. The Svetlichny operator is defined by
\begin{eqnarray}\label{zs1}& &\mathbb{S}=A_0(B_0+B_1)C_0+A_0(B_0-B_1)C_1\nonumber\\& &+A_1(B_0-B_1)C_0-A_1(B_0+B_1)C_1.
\end{eqnarray}
If a state $\rho$ admits bi-local hidden variable model, then the
expectation value of the Svetlichny operator satisfies the following
Svetlichny's inequality,
\begin{equation}\label{zs2}
\mathbb{S}(\rho)=|Tr(\mathbb{S}\rho)|\leq 4.
\end{equation}
If the inequality (\ref{zs2}) is violated, $\rho$ must be a genuine three-qubit
nonlocally correlated state.

In 1974, Hawking discovered black hole evaporation caused by thermal
radiation emitted by the black hole \cite{swha}. Soon after, Unruh
in 1976 proposed a very important prediction based on the effect of
Hawking radiation, which says that a vacuum state observed by an
observer who stays in flat Minkowski spacetime would be detected as
a thermal state from another observer who hovers near the event
horizon of the black hole with uniform acceleration \cite{wgun,lcag}.
In recent years, as the intersection of classical informatics,
quantum mechanics, special and general relativity and quantum field
theory, the problem of quantum information in non-inertial system
and curved spacetime has become a new hotspot in the research of
quantum information science \cite{ifrb,pmir,jqsj,eljj,jqja,jqjb,emlj,jwjja,skmk,jwjjb,bnmm,rbtc,ztjja,
sxjl,ztjjb,skjj,ydzy,nfri,xzjj,wgqs,aqjw,smhsa,smhsb,smzh,smhsc,lfxl,smht}.
Undoubtedly, the study of quantum information under the framework of
relativity, especially in the non-inertial frame and curved
spacetime, not only is of great significance to the
development and perfectness of quantum information theory, but also
plays a positive role in understanding the entropy and information
paradox in black holes. For examples, Fuentes et
al. \cite{ifrb} show that the entanglement in noninertial frames
is characterized by the observer-dependent property.
Mart\'in-Mart\'inez et al. \cite{eljj} studied the entanglement
degradation affected by the Hawking effect of Schwarzschild black
hole. Mann et al. \cite{rbtc} pointed out that incorporating concepts
from quantum theory into relativity can yield novel and interesting
effects. More recently, Wu et al. \cite{smhsc} studied the GTN and
the genuine tripartite
entanglement (GTE) of Dirac fields in the background of a Schwarzschild black
hole. Li et al. \cite{lfxl} explored the tripartite entropic
uncertainty and genuine tripartite quantum coherence of Dirac fields
in the background of the Garfinkle-Horowitz-Strominger (GHS)
dilation space-time.

March attention has been paid to the multipartite
Greenberger-Horne-Zeilinger (GHZ) state
$|GHZ\rangle=\frac{1}{\sqrt{2}}(|000\rangle+|111\rangle)$ or the
generalized GHZ states under the Hawking radiation. It is recently
found in Refs. \cite{smhsc} and \cite{lfxl} that the GTN is not
redistributable, but GTE can be redistributed through Hawking
effect. With the growth of the Hawking temperature, the physically
accessible GTN decreases, but physically inaccessible GTN is not
generated. The GTE however behaves differently: with the loss of the
physically accessible GTE, the physically inaccessible GTE is
generated continually through Hawking effect. This is inconsistent
with intuition since quantum entanglement and quantum nonlocality
are in consistent for pure states \cite{nsin,sqcc,szha}. On the
other hand, previous studies have shown that for some three-qubit
pure states (GGHZ state, MS state and so on), it has been observed
only that the GTN decreases \cite{asrm,ztjw}. But the
measurement-induced-nonlocality of two-qubit mixed states can
increase via the Unruh effect or near a black hole
\cite{ztjjc,azds}. To clarify such problems, in this article we
study the GTN of multipartite mixed states for Dirac fields in the
background of a Schwarzschild black hole. The paper is organized as
follows. In the second section, we briefly recall the quantization
of Dirac fields in the background of Schwarzschild black hole, and
the GTN, GTE and quantum coherence for three-qubit X-states. In the
third section, we calculate and analyze the GTN of a detailed mixed
state after one or bipartite acceleration. A novel phenomenon that
is basically different from that of pure states is illustrated: the
Hawking effect may not only degrades the physically accessible GTN,
but also enhances the physically inaccessible GTN. The ``sudden
death" of physically accessible GTN takes place at some critical
Hawking temperature, while ``sudden birth"  happens to the
physically inaccessible GTN. We show that such phenomenon may be not
observed for other mixed states. In the fourth section, we consider
the quantum correlation GTE and quantum coherence. Analytical
tradeoff relations among the distributions of GTE and coherence are
derived. The last section is the conclusions and discussions.

\section{Quantization of Dirac fields in a Schwarzschild black
hole}

We first recall the evolution process of vacuum
state of Dirac fields in a Schwarzschild black hole \cite{smhsc,lfxl}.
The Dirac equation under a general background spacetime can be
written as \cite{drja},
\begin{eqnarray}\label{zs6}
[\gamma^ae_a^{\mu}(\partial_{\mu+\Gamma_{\mu}})]\Phi=0,
\end{eqnarray}
where$\gamma^a$ are the Dirac matrices, $e_a^{\mu}$
is the inverse of the tetrad $e_{\mu}^a$,
$\Gamma_{\mu}=\frac{1}{8}[\gamma^a,\gamma^b]e_a^{\nu}e_{b\nu;\mu}$
are the spin connection coefficients. The metric of the
Schwarzschild black hole can be written as \cite{dgta,jqjb}
\begin{eqnarray}
ds^2=&-&(\frac{r-2M}{r-2D})dt^2+(\frac{r-2M}{r-2D})^{-1}dr^2
\nonumber\\ &+& r(r-2D)(d\theta^2+\sin^2\theta
d\phi^2),
\end{eqnarray}
where $M$ denotes the mass of the black
hole, and $D$ denotes the parameter of the dilation field. The
Hawking temperature is given by $T=\frac{1}{8\pi(M-D)}$.
The thermal Fermi-Dirac distribution of particles with Hawking temperature
$T$ was observed in Refs. \cite{stgw,tdrr}.

Solving the Dirac equation (\ref{zs6}) near the event horizon
of black hole, one gets a set of positive frequency outgoing solutions for
the inside and outside regions of the event horizon,
\cite{eljj,jqja} \begin{eqnarray}\label{zs7}
\phi_k^{I+}=\xi
e^{-i\omega\mu}, \ \ \phi_k^{II+}=\xi e^{i\omega\mu},
\end{eqnarray}
where $k$ represents the field mode, $\xi$ denotes four-component
Dirac spinor, $\omega$ is a monochromatic frequency and
$\mu=t-r_{\ast}$, where the tortoise
coordinate $r_{\ast}=r+2M\ln\frac{r-2M}{2M}$. $\phi_k^{I+}$
and $\phi_k^{II+}$ are usually called Schwarzschild modes.
Particles and antiparticles are classified according
to future-directed timelike Killing vector under each region.

An analytic continuation of (\ref{zs7}) \cite{lfxl} gives rise to a complete basis of positive
energy modes, i.e., the Kruskal modes \cite{tdrr}. Then one can use
the Schwarzschild mode and the Kruskal mode to expand the Dirac field,
respectively, leading to the Bogoliubov transformations between
annihilation operator and creation operator under the Schwarzschild
and Kruskal coordinates \cite{smzh}. After properly normalizing the
state vector, the vacuum state and excited state of the Kruskal
particle in the single-mode approximation are given by
\begin{eqnarray}\label{zs8}
&|0\rangle=\cos\beta|0\rangle_{I}^{+}|0\rangle_{II}^{-}
+\sin\beta|1\rangle_{I}^{+}|1\rangle_{II}^{-},\\
\label{zs9}
&|1\rangle=|1\rangle_{I}^{+}|0\rangle_{II}^{-},
\end{eqnarray}
where $|n\rangle_{I}$ and $|n\rangle_{II}$ are the number states for
the particle outside the region (physically accessible) and the
antiparticle inside the region of the event horizon (physically
inaccessible), respectively, the superscripts $+$ and $-$ indicate
the particle and antiparticle, respectively,
$\cos\beta=(e^{-\frac{\omega}{T}})^{-\frac{1}{2}}$ and
$\sin\beta=(e^{\frac{\omega}{T}})^{-\frac{1}{2}}$.

In the following we will consider three-qubit X-state give by density matrix,
$$
\rho_X=\left(\begin{array}{cccccccc}
    d_1 &  0 & 0 & 0 &  0 & 0 & 0 & f_1 \\
    0 &  d_2 & 0 & 0 &  0 & 0 & f_2 & 0 \\
    0 &  0 & d_3 & 0 &  0 & f_3 & 0 & 0 \\
    0 &  0 & 0 & d_4 &  f_4 & 0 & 0 & 0 \\
    0 &  0 & 0 & f_4^{\ast} &  e_4 & 0 & 0 & 0 \\
    0 &  0 & f_3^{\ast} & 0 &  0 & e_3 & 0 & 0 \\
    0 &  f_2^{\ast} & 0 & 0 &  0 & 0 & e_2 & 0 \\
    f_1^{\ast} &  0 & 0 & 0 &  0 & 0 & 0 & e_1
\end{array}\right)
$$
in the orthogonal basis $\{ |000\rangle,|001\rangle,\cdots,|111\rangle\}$.
The GTN (Svetlichny value) of $\rho_X$ is simply given by \cite{kwzj}
\begin{eqnarray}\label{zs3}
S(\rho_X)=\max\{8\sqrt{2}|f_i|,4|N|\},
\end{eqnarray}
where $N=d_1-d_2-d_3+d_4-e_4+e_3+e_2-e_1$.
The GTE of $\rho_X$ is given by \cite{zzsm}
\begin{eqnarray}\label{zs4}
E(\rho_X)=2\max\{0,|f_i|-m_i\}, i=1,2,3,4,
\end{eqnarray}
where $m_i=\sum_{j\neq i}^4\sqrt{d_je_j}$.
The quantum coherence of $\rho_X$ is given by \cite{taml}
\begin{eqnarray}\label{zs5}
C(\rho_X)=C_{l_1}(\rho_X)=\sum_{i\neq j}|\rho_{ij}|=2\sum_{i=1}^4|f_i|.
\end{eqnarray}

\section{Evolution of GTN for mixed states in Schwarzschild black hole}
We first consider the initial tripartite mixed state $\rho_1=p|
GHZ\rangle\langle GHZ|+(1-p)| 000\rangle\langle 000|$ of the dirac
fields shared by Alice, Bob and Charlie in the asymptotically flat
region, where $p\in[0,1]$.

{\bf Case(1): Alice and Bob stay at an asymptotically flat region,
while Charlie hovers outside the event horizon of the black hole.}
Applying the transformation of Eqs. (\ref{zs8}) and (\ref{zs9}) for
detector C, we obtain a 4-partite state $\rho_{ABC_IC_{II}}$
associated with subsystems $A$, $B$ and $C_I$ observed by Alice, Bob
and Charlie outside the event horizon of black hole, respectively,
and the subsystem $C_{II}$ observed by anti-Charlie inside the event
horizon. Since the interior region of black hole is causally
disconnected from the exterior region, Alice, Bob and Charlie cannot
access the modes inside the event horizon. The mode $C_I$ outside
the event horizon is called the accessible modes, and the mode
$C_{II}$ inside the event horizon the inaccessible modes. By tracing
over the inaccessible mode on $\rho_{ABC_IC_{II}}$, we obtain the
reduced density matrix,
\begin{eqnarray}
& & \rho_{ABC_I} =  \nonumber\\ & &\left(\begin{array}{cccccccc}
    (\frac{2-p}{2})\cos^2 r_c &  0 & 0 & 0 &  0 & 0 & 0 & \frac{p}{2}\cos r_c \\
    0 &  (\frac{2-p}{2})\sin^2 r_c & 0 & 0 &  0 & 0 & 0 & 0 \\
    0 &  0 & 0 & 0 &  0 & 0 & 0 & 0 \\
    0 &  0 & 0 & 0 &  0 & 0 & 0 & 0 \\
    0 &  0 & 0 & 0 &  0 & 0 & 0 & 0 \\
    0 &  0 & 0 & 0 &  0 & 0 & 0 & 0 \\
    0 &  0 & 0 & 0 &  0 & 0 & 0 & 0 \\
    \frac{p}{2}\cos r_c  &  0 & 0 & 0 &  0 & 0 & 0 & \frac{p}{2}
\end{array}\right)\nonumber,
\end{eqnarray}
where $\cos r_c=(e^{-\frac{\omega}{T_c}})^{-\frac{1}{2}}$ and $\sin
r_c=(e^{\frac{\omega}{T_c}})^{-\frac{1}{2}}$ are functions of the
Hawking temperature $T_c$. For simplicity, we set
$\omega=1$ and instead of $T_c$, we use $r_c$ in the following analysis.

From Eq. (\ref{zs3}), we obtain the GTN of the state $\rho_{ABC_I}$,
$$
S(\rho_{ABC_I})=\max\{8\sqrt{2}|\frac{p}{2}\cos r_c |,4|(1-\frac{p}{2})\cos2r_c-\frac{p}{2}|\}.
$$
In FIG. 1 (a) we plot the Svetlichny value
$S(\rho_{ABC_I})$ as function of $r_c$ (the Hawking temperature $T$)
and $p$. We find that, with the increase of Hawking temperature
$T$ and the decreasing of parameter $p$ from $1$ to $0$,
$S(\rho_{ABC_I})$ is larger than $4$ at first and then smaller than $4$.
This implies two interesting aspects. One is that the thermal noise introduced by
Hawking temperature destroys the physically accessible GTN between
Alice, Bob and Charlie. And the ``sudden death" takes place at a
critical temperature $T_c$. Another phenomenon is that with the
decrease of the parameter $p$, the GTN transition from the maximally
entangled initial state to the complete separable one is
also influenced by Hawking effect.
\begin{figure}[ptb]
\includegraphics[width=0.45\textwidth]{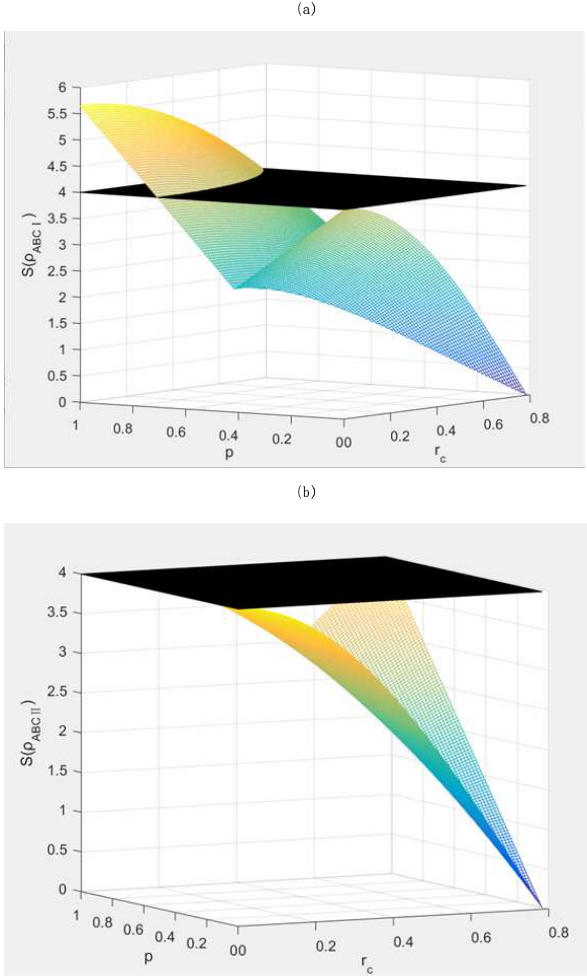}
\caption{ {\bf Case(1): Alice and Bob stay at an asymptotically flat
region, while Charlie hovers outside the event horizon of the black
hole.} (a)  the Svetlichny value $S(\rho_{ABC_I})$ as function of
$r_c$ (the Hawking temperature $T$) and $p$.  (b) the Svetlichny
value $S(\rho_{ABC_{II}})$ as function of $r_c$  and $p$. }
\end{figure}

Similarly, by tracing over the accessible mode $C_I$ in
$\rho_{ABC_IC_{II}}$ we get the reduced density matrix,
\begin{eqnarray}& & \rho_{ABC_{II}} =  \nonumber\\ &
&\left(\begin{array}{cccccccc}
    (\frac{2-p}{2})\cos^2 r_c &  0 & 0 & 0 &  0 & 0 & 0 & 0 \\
    0 &  (\frac{2-p}{2})\sin^2 r_c & 0 & 0 &  0 & 0 & \frac{p}{2}\sin r_c & 0 \\
    0 &  0 & 0 & 0 &  0 & 0 & 0 & 0 \\
    0 &  0 & 0 & 0 &  0 & 0 & 0 & 0 \\
    0 &  0 & 0 & 0 &  0 & 0 & 0 & 0 \\
    0 &  0 & 0 & 0 &  0 & 0 & 0 & 0 \\
    0 &  \frac{p}{2}\sin r_c & 0 & 0 &  0 & 0 & \frac{p}{2}& 0 \\
    0 &  0 & 0 & 0 &  0 & 0 & 0 & 0
\end{array}\right)\nonumber
\end{eqnarray}
and the GTN of the state $\rho_{ABC_{II}}$,
$$
S(\rho_{ABC_{II}})=\max\{8\sqrt{2}|\frac{p}{2}\sin r_c |,4|(\frac{2-p}{2})\cos2r_c+\frac{p}{2}|\}.
$$
From FIG. 1 (b) we see that
$S(\rho_{ABC_{II}})$ is always small than $4$ for any $r_c$ and $p$.
Thus the physically inaccessible GTN between modes $A,B$ and
$C_{II}$ cannot be produced.

Case(2): Alice still stays at an asymptotically
flat region, while Bob and Charlie hover outside the event horizon
of the black hole. Then in terms of the Kruskal
modes for Alice and Schwarzschild modes for Bob and Charlie,
the initial state $\rho_1$ can be rewritten as a
5-partite entangled state $\rho_{AB_{I}B_{II}C_{I}C_{II}}$ consisted
by subsystem $A$ observed by Alice, subsystems $B_I$ and
$C_I$ observed by Bob and Charlie outside the event horizon of black
hole, and subsystems $B_{II}$ and $C_{II}$ observed by anti-Bob and
anti-Charlie inside the event horizon, respectively. We call
the modes $B_I$ and $C_I$ outside the event horizon the accessible
modes, and the modes $B_{II}$ and $C_{II}$ inside the event horizon
the inaccessible modes. Tracing over the inaccessible modes in
state $\rho_{AB_{I}B_{II}C_{I}C_{II}}$, we obtain the reduced
density operator,
\begin{eqnarray}& & \rho_{AB_IC_{I}} = \left(\begin{array}{cccccccc}
    d_1 &  0 & 0 & 0 &  0 & 0 & 0 &  f_1\\
    0 &  d_2 & 0 & 0 &  0 & 0 & 0 & 0 \\
    0 &  0 & d_3 & 0 &  0 & 0 & 0 & 0 \\
    0 &  0 & 0 & d_4 &  0 & 0 & 0 & 0 \\
    0 &  0 & 0 & 0 &  0 & 0 & 0 & 0 \\
    0 &  0 & 0 & 0 &  0 & 0 & 0 & 0 \\
    0 &  0 & 0 & 0 &  0 & 0 & 0 & 0 \\
    f_1 &  0 & 0 & 0 &  0 & 0 & 0 & \frac{p}{2}
\end{array}\right)\nonumber,
\end{eqnarray}
where $d_1=(\frac{2-p}{2})\cos^2 r_b
\cos^2 r_c,$ $d_2=(\frac{2-p}{2})\cos^2 r_b \sin^2 r_c,$
$d_3=(\frac{2-p}{2})\sin^2 r_b \cos^2 r_c,$
$d_4=(\frac{2-p}{2})\sin^2 r_b \sin^2 r_c$ and $f_1=\frac{p}{2}\cos r_b
\cos r_c.$ Here, $\cos r_b=(e^{-\frac{\omega}{T_b}})^{-\frac{1}{2}}$
and $\sin r_b=(e^{\frac{\omega}{T_b}})^{-\frac{1}{2}}$ are functions
of the Hawking temperature $T_b$.
We obtain the GTN for $\rho_{AB_IC_{I}}$,
\begin{eqnarray}& & S(\rho_{AB_IC_{I}}) = \nonumber\\ &
&\max\{4\sqrt{2}p|\cos r_b\cos r_c
|,4|(\frac{2-p}{2})\cos2r_b\cos2r_c-\frac{p}{2}|\}.\nonumber
\end{eqnarray}

In FIG. 2, we plot the Svetlichny value $S(\rho_{AB_IC_I})$ as
function of $r_b=r_c$ and $p$. I it shown that
$S(\rho_{AB_IC_I})$ is larger than $4$ at first and then smaller
than $4$ with the increase of Hawking temperature $T$ and the
decrease of $p$. This implies that the thermal
noise introduced by Hawking temperature destroys the physically
accessible GTN between Alice, Bob and Charlie. The
``sudden death" of such physically accessible GTN takes place at some critical temperature $T_b$.
\begin{figure}[ptb]
\includegraphics[width=0.45\textwidth]{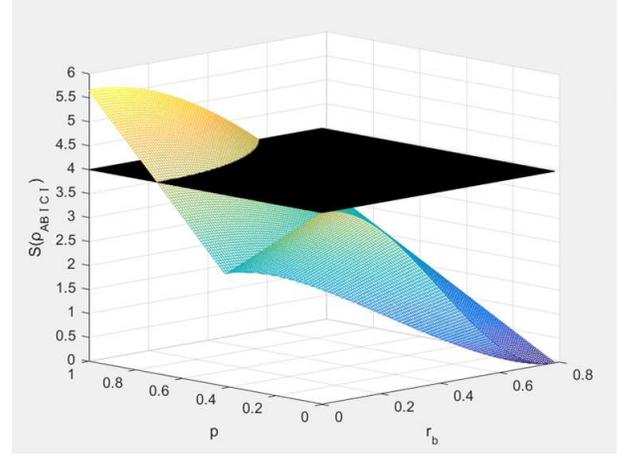}\caption{ Case(2): The Svetlichny value $S(\rho_{AB_IC_I})$ as a function of $r_c=r_b$ and $p$ for
the case that {\bf Alice still stays at an asymptotically flat
region, while Bob and Charlie hover outside the event horizon of the
black hole.}}
\end{figure}

The reduced density matrix $\rho_{AB_IB_{II}}$ can be
expressed as
\begin{eqnarray}& & \rho_{AB_IB_{II}} =
\left(\begin{array}{cccccccc}
    d_1 &  0 & 0 & 0 &  0 & 0 & 0 &  f_1\\
    0 &  0 & 0 & 0 &  0 & 0 & 0 & 0 \\
    0 &  0 & 0 & 0 &  0 & 0 & 0 & 0 \\
    0 &  0 & 0 & 0 &  0 & 0 & 0 & 0 \\
    0 &  0 & 0 & 0 &  0 & 0 & 0 & 0 \\
    0 &  0 & 0 & 0 &  0 & 0 & 0 & 0 \\
    0 &  0 & 0 & 0 &  0 & 0 & \frac{p}{2} & 0 \\
    f_1 &  0 & 0 & 0 &  0 & 0 & 0 & e_1
\end{array}\right)\nonumber,
\end{eqnarray}
with the matrix elements give by $d_1=(\frac{2-p}{2})\cos^2 r_b,$
$f_1=(\frac{2-p}{2})\cos r_b \sin r_b$ and
$e_1=(\frac{2-p}{2})\sin^2 r_b$. In FIG. 3, we plot the Svetlichny
value $S(\rho_{AB_IB_{II}})$ as function of $r_b=r_c$ and $p$. We
find that $S(\rho_{AB_IB_{II}})$ is smaller than $4$ at first and
then larger than $4$ with the increase of Hawking temperature $T$
and the decrease of $p$. This phenomenon is quite novel as the birth
of the physically inaccessible GTN for $\rho_{AB_IB_{II}}$ has never
been observed in previous literature with pure initial states
\cite{smhsc,lfxl}. To see this new phenomenon more clearly, we draw
$S(\rho_{AB_IB_{II}})$ corresponding to different values of $p$ in
FIG. 3(b). We see that when $p=1$, the birth of GTN does not show
up. As the value of $p$ becomes smaller, GTN becomes more than $4$
when $p=\frac{1}{2},\frac{1}{4}$ and so on. This result shows that
the GTN can pass through the event horizon of black hole for the
case of mixed states.

Additionally, we note that in FIG 3 (b), the phenomenon of GTN can
also be observed for $p=0$, namely, the original state is separable.
This is impossible in inertial fields, but it indeed happens in
non-inertial fields with acceleration and rotation. Similar
phenomenon has been observed for quantum entanglement
\cite{mmmd,msgm}, where entangled states are successfully obtained
under rotation effect when the initial state is separable. These
facts also show the peculiarity of quantum information in the
framework of relativity.

\begin{figure}[ptb]
\includegraphics[width=0.45\textwidth]{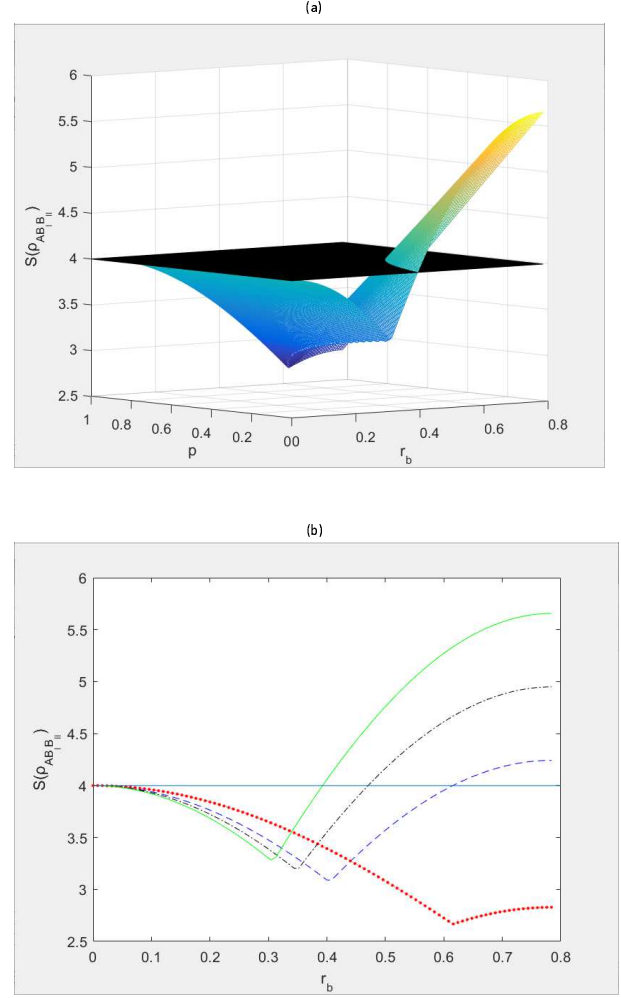}\caption{
{\bf Case(2): Alice still stays at an asymptotically flat region,
while Bob and Charlie hover outside the event horizon of the black
hole.} (a) the Svetlichny value $S(\rho_{AB_IB_{II}})$ as function
of $r_b$ and $p$. (b) the Svetlichny value $S(\rho_{AB_IB_{II}})$ as
function of $r_b$. The
 dotted red, the dashed blue, the dot-dashed gray and the solid
green lines are for $p=1,\frac{1}{2},\frac{1}{4},0$, respectively.}
\end{figure}

Due the symmetry under the interchange of Bob and Charlie, we have
$S(\rho_{AC_IC_{II}})=S(\rho_{AB_IB_{II}})$ and
$S(\rho_{AB_IC_{II}})=S(\rho_{AB_{II}C_I})$.
Therefore, the GTN of $\rho_{AC_IC_{II}}$ has similar properties to
$\rho_{AB_IB_{II}}$ under the influence of Hawking effect. We
only need to consider the GTN for the rest two types of reduced
states $\rho_{AB_IC_{II}}$ and $\rho_{AB_{II}C_{II}}$ of
$\rho_{AB_{I}B_{II}C_{I}C_{II}}$. By direct calculation, we have
\begin{eqnarray}& & \rho_{AB_IC_{II}} =
\left(\begin{array}{cccccccc}
    d_1 &  0 & 0 & 0 &  0 & 0 & 0 & 0 \\
    0 &  d_2 & 0 & 0 &  0 & 0 & f_2 & 0 \\
    0 &  0 & d_3 & 0 &  0 & 0 & 0 & 0 \\
    0 &  0 & 0 & d_4 &  0 & 0 & 0 & 0 \\
    0 &  0 & 0 & 0 &  0 & 0 & 0 & 0 \\
    0 &  0 & 0 & 0 &  0 & 0 & 0 & 0 \\
    0 &  f_2 & 0 & 0 &  0 & 0 & \frac{p}{2} & 0 \\
    0 &  0 & 0 & 0 &  0 & 0 & 0 & 0
\end{array}\right)\nonumber,
\end{eqnarray}
where $d_1=(\frac{2-p}{2})\cos^2 r_b
\cos^2 r_c,$ $d_2=(\frac{2-p}{2})\cos^2 r_b \sin^2 r_c,$
$d_3=(\frac{2-p}{2})\sin^2 r_b \cos^2 r_c,$
$d_4=(\frac{2-p}{2})\sin^2 r_b \sin^2 r_c$ and $f_2=\frac{p}{2}\cos r_b
\cos r_c$,
and
\begin{eqnarray}& & \rho_{AB_{II}C_{II}} =
\left(\begin{array}{cccccccc}
    d_1 &  0 & 0 & 0 &  0 & 0 & 0 & 0 \\
    0 &  d_2 & 0 & 0 &  0 & 0 & 0 & 0 \\
    0 &  0 & d_3 & 0 &  0 & 0 & 0 & 0 \\
    0 &  0 & 0 & d_4 &  f_4 & 0 & 0 & 0 \\
    0 &  0 & 0 & f_4 &  \frac{p}{2} & 0 & 0 & 0 \\
    0 &  0 & 0 & 0 &  0 & 0 & 0 & 0 \\
    0 &  0 & 0 & 0 &  0 & 0 & 0 & 0 \\
    0 &  0 & 0 & 0 &  0 & 0 & 0 & 0
\end{array}\right)\nonumber,
\end{eqnarray}
where $d_1=(\frac{2-p}{2})\cos^2 r_b
\cos^2 r_c,$ $d_2=(\frac{2-p}{2})\cos^2 r_b \sin^2 r_c,$
$d_3=(\frac{2-p}{2})\sin^2 r_b \cos^2 r_c,$
$d_4=(\frac{2-p}{2})\sin^2 r_b \sin^2 r_c$ and $f_4=\frac{p}{2}\sin r_b
\sin r_c.$

Using (\ref{zs3}) we obtain $S(\rho_{AB_IC_{II}})$ and
$S(\rho_{AB_{II}C_{II}})$, respectively. FIG. 4
shows that the GTN cannot be generated between $A, B_I $ and $C_{II}$ or
between $A, B_{II}$ and $ C_{II}$.
\begin{figure}[ptb]
\includegraphics[width=0.45\textwidth]{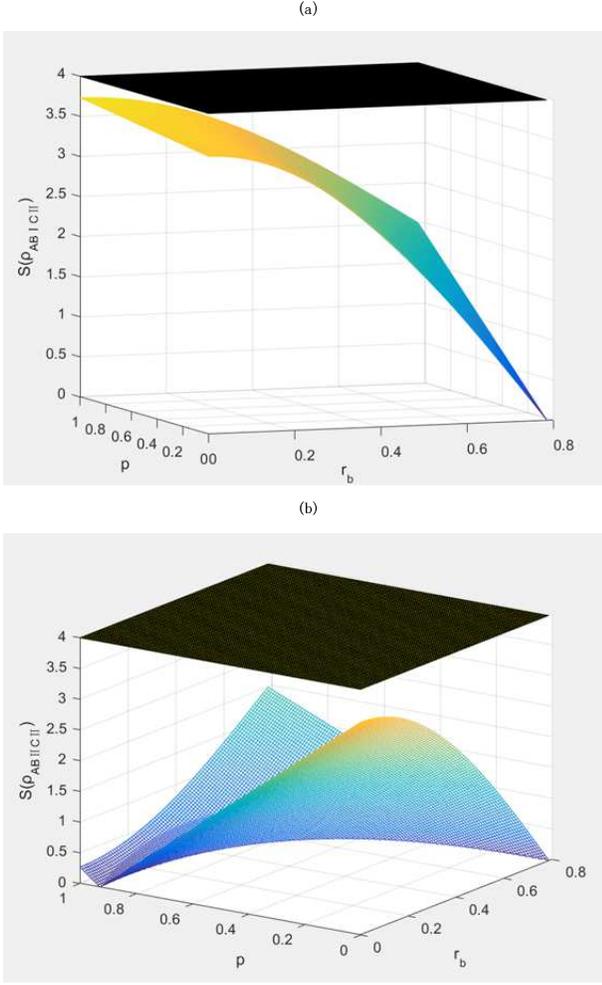}\caption{
{\bf Case(2): Alice still stays at an asymptotically flat region,
while Bob and Charlie hover outside the event horizon of the black
hole.} (a) the Svetlichny value $S(\rho_{AB_IC_{II}})$ as function
of $r_c=r_b$ and $p$. (b) The Svetlichny value
$S(\rho_{AB_{II}C_{I}})$ as function of $r_c=r_b$ and $p$.}
\end{figure}

We have used the initial mixed state  $\rho_1=p|
GHZ\rangle\langle GHZ|+(1-p)| 000\rangle\langle 000|$ to display the
novel phenomenon of sudden birth of GTN. However, such phenomenon
may not appear for other mixed states. Let us consider the mixed state
$\rho_2=p| GHZ\rangle\langle GHZ|+(1-p)| 111\rangle\langle 111|$. For
$\rho_2$ the GTN for other reduced states are similar to $\rho_1$, except for
$S(\rho_{AB_IB_{II}})$ which is always less than $4$, namely, no
GTN is generated, See FIG. 5.
The only difference between the states $\rho_1$ and $\rho_2$
is that the vector $|000\rangle$ is replaced with $|111\rangle$.
Their different behaviors on the new generation of physically inaccessible
GTN are due to that the evolution process of vacuum state of Dirac fields in a
Schwarzschild black hole has different effects on $|0\rangle$ and
$|1\rangle$.
\begin{figure}[ptb]
\includegraphics[width=0.45\textwidth]{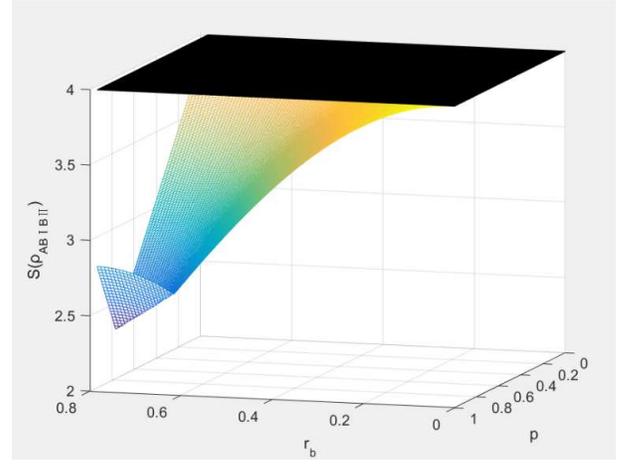}\caption{
{\bf Case(2): Alice still stays at an asymptotically flat region,
while Bob and Charlie hover outside the event horizon of the black
hole.} The Svetlichny value $S(\rho_{AB_IB_{II}})$ with respect to
the initial state $\rho_2$ as function of $r_b$ and $p$.}
\end{figure}

\section{Evolution of the GTE and quantum coherence in Schwarzschild
black hole}

In this section, we explore the GTE
and quantum coherence under evolution of $\rho_1$
after the acceleration effect for one or two
accelerated detector(s) coupled to the vacuum field.

It is very coincidental that according to Eqs. (\ref{zs4}) and
(\ref{zs5}), we find that for all the tripartite reduced states
in the previous section, the GTE is equal to the
quantum coherence. For the case (1), we have
$$
\begin{array}{l}
E(\rho_{ABC_I})=C(\rho_{ABC_I})=p\cos r_c,\\
E(\rho_{ABC_{II}})=C(\rho_{ABC_{II}})=p\sin r_c.
\end{array}
$$
For the case (2), we obtain
$$
\begin{array}{l}
E(\rho_{AB_IC_I})=C(\rho_{AB_IC_I})=p|\cos r_b\cos r_c|,\\
E(\rho_{AB_IC_{II}})=C(\rho_{AB_IC_{II}})=p|\cos r_b\sin r_c|,\\
E(\rho_{AB_IB_{II}})=C(\rho_{AB_IB_{II}})=(2-p)|\cos r_b\sin r_b|,\\
E(\rho_{AB_{II}C_{II}})=C(\rho_{AB_{II}C_{II}})=p|\sin r_b\sin r_c|.
\end{array}
$$
Due to the symmetry between Bob and Charlie, it is easy to find that
$$
\begin{array}{l}
E(\rho_{AC_{I}C_{II}})=E(\rho_{AB_{I}B_{II}})=(2-P)|\sin r_c\cos r_c|,\\
E(\rho_{AB_{II}C_{I}})=E(\rho_{AB_IC_{II}})=p|\cos r_c\sin r_b|.
\end{array}
$$

The tradeoff relationship and the monogamy relations are important
issues in characterizing the quantum correlations in quantum information
theory, which are deeply related to quantum uncertainty and incompatibility
principle. Noted that the acceleration of Bob and Charlie is completed under the same Hawking
radiation. Hence, the corresponding changes of the Hawking temperatures
$T_b$ and $T_c$ are the same. Therefore, the functions $\cos r_b$ and
$\cos r_c$ corresponding to $T$ are independent individually, namely,
$\cos r_b=\cos r_c$ and $\sin r_b=\sin r_c$. Therefore, we have the following
tradeoff relations between the physically accessible
GTE (quantum coherence) and the physically inaccessible GTE (quantum
coherence),
\begin{eqnarray}
&E^2(\rho_{ABC_I})+E^2(\rho_{ABC_{II}})=p^2=E^2(\rho_1),\\
&E(\rho_{AB_IC_I})+E(\rho_{AB_{II}C_{II}})=p=E(\rho_1),\label{zs11}\\
&E^2(\rho_{AB_IC_I})+E^2(\rho_{AB_{II}C_I})+E^2(\rho_{AB_IC_{II}})\nonumber\\
&+E^2(\rho_{AB_{II}C_{II}})=p^2. \label{zs12}
\end{eqnarray}
The above trade-off relations give restrictions on the
redistribution of entanglement from physically accessible to
physically inaccessible patterns. For instance, (\ref{zs11}) shows
that the total physically accessible GTE
$E(\rho_{AB_IC_I})$ and the physically inaccessible GTE
$E(\rho_{AB_{II}C_{II}})$ is equal to the initial GTE of $\rho_1$.
Therefore, under the time evolution how much the physically
accessible GTE decreases implies how much the physically inaccessible GTE
increases. The Eq. (\ref{zs12}) poses a strong restrictive
relationship among the four genuine tripartite entanglement. These tradeoff
relations shed new lights on understanding the quantum
information theory in the framework of relativity.

\section{Conclusions and discussions}
Quantum nonlocality is a distinctive feature of quantum mechanics.
We have investigated the effect of Hawking radiation on
GTN for Dirac fields in Schwarzschild spacetime.
We first time show that the Hawking effect can not only degrades the physically
accessible GTN, but also enhance the physically
inaccessible GTN. The ``sudden death" of the physically
accessible GTN at certain critical Hawking temperature and the
``sudden birth" of the physically
inaccessible GTN imply that, with the loss of the physically
accessible GTN, the physically inaccessible GTN is generated
continually through Hawking effect. This phenomenon is observed first time
for GTN. If we regard the redistribution of nonlocality as a kind of
phenomenon of information tunnelling, this phenomenon shows
that not only the flow of quantum entanglement can pass through the
black hole perspective, but also the nonlocality can either.
This phenomenon also enriches further
understanding of relativistic quantum information.

We have also investigated the problem for quantum correlation GTE and
quantum coherence. Interesting trade-off relations are derived among
the GTEs and the quantum coherence, which restrict the distribution
of GTE and coherence in the subsystems during the evolution.
With regard to the related theories of entanglement measurement under the effect of Hawking radiation for Dirac fields in Schwarzschild spacetime, the previous literatures mainly
focused on pure initial states. Based on initial mixed states, we have shown
the new phenomenon that the physically inaccessible
GTN is generated continually through Hawking effect. From the view of resource theory in quantum mechanics, the quantum correlations including entanglement and nonlocality, as well as the coherence may have universal redistribution properties. Our results may highlight new understanding of relativistic quantum information.

\bigskip
\section*{Acknowledgments}
This work is supported by the Hainan Provincial Natural Science
Foundation of China under Grant Nos.121RC539; the National Natural
Science Foundation of China (NSFC) under Grant Nos 12075159 and
12171044; the specific research fund of the Innovation Platform for
Academicians of Hainan Province under Grant No. YSPTZX202215;
Beijing Natural Science Foundation (Grant No. Z190005).

\end{document}